\documentclass[pre,onecolumn,notitlepage,showpacs,showkeys,superscriptaddress,preprintnumbers,floatfix]{revtex4-1}

\usepackage{etex}
\usepackage{ifpdf}
\usepackage{hyperref}
\usepackage{dcolumn}
\usepackage{url}
\usepackage{amsmath}
\usepackage{amscd}
\usepackage{amsfonts}
\usepackage{amssymb}
\usepackage{amsthm}
\usepackage{xfrac}
\usepackage{braket}
\usepackage{bm}   % bold math
\usepackage{bbm}
\usepackage{verbatim}
\usepackage{mathtools}
\usepackage{stmaryrd}
\usepackage{xcolor}
\usepackage{setspace}
\usepackage{tikz}
\usetikzlibrary{arrows}
\usetikzlibrary{automata}
\usetikzlibrary{calc}
\usetikzlibrary{decorations.pathreplacing}
\usetikzlibrary{patterns}
\usetikzlibrary{positioning}
\usetikzlibrary{plotmarks}
\usetikzlibrary{shapes}

\usepackage{pgfplots}
\pgfplotsset{compat=newest}

\tikzstyle{vaucanson}=[
  node distance=2.5cm,
  bend angle=15,
  auto,
  every loop/.style={},
  every edge/.style={->,draw=black,line width=1.2,>=latex,shorten <=1pt,shorten >=1pt},
  every state/.style={draw=black,line width=2,font=\large},
  loop right/.style={right,out=22,in=-22,loop},
  loop above/.style={above,out=112,in=68,loop},
  loop left/.style={left,out=202,in=158,loop},
  loop below/.style={below,out=292,in=248,loop},
  loop above right/.style={right,out=67,in=23,loop},
  loop above left/.style={left,out=157,in=113,loop},
  loop below left/.style={left,out=247,in=203,loop},
  loop below right/.style={right,out=337,in=293,loop},
]

\theoremstyle{plain}    
\theoremstyle{plain}    
\theoremstyle{plain}    
\theoremstyle{plain}    
\theoremstyle{plain}    
\theoremstyle{plain}    
\theoremstyle{plain}    \newtheorem{The}{Theorem}
\theoremstyle{plain}    
\theoremstyle{plain}    \newtheorem*{ProThe}{Proof}
\theoremstyle{plain}    \newtheorem{Prop}{Proposition}
\theoremstyle{plain}    
\theoremstyle{plain}    \newtheorem*{ProProp}{Proof}
\theoremstyle{plain}    
\theoremstyle{plain}    
\theoremstyle{plain}    
\theoremstyle{plain}    
\theoremstyle{plain}    

\newcommand{\eM}     {\mbox{$\epsilon$-machine}}
\newcommand{\eMs}    {\mbox{$\epsilon$-machines}}

\newcommand{\Gen}{\mathcal{G}}
\newcommand{\gen}{g}

\newcommand{\MS}{X}
\newcommand{\MT}{\mathcal{T}}
\newcommand{\ms}{x}
\newcommand{\mt}{\tau}

\newcommand{\MeasAlphabet}  {\mathcal{A}}
\newcommand{\Past} { \smash{\overleftarrow {(\MS,\MT)}} }
\newcommand{\past} { \smash{\overleftarrow {(\ms,\mt)}} }

\newcommand{\pastprime} { {\past}^{\prime}}
\newcommand{\Future}   { \smash{\overrightarrow{(\MS,\MT)}} }
\newcommand{\future}   { \smash{\overrightarrow{(\ms,\mt)}} }
\newcommand{\CausalState}   { \mathcal{S} }

\newcommand{\causalstate}   { \sigma }
\newcommand{\CausalStateSet}    { \boldsymbol{\CausalState} }
\newcommand{\AlternateState}    { \mathcal{R} }

\newcommand{\AlternateStateSet} { \boldsymbol{\AlternateState} }

\newcommand{\Prob}      {\Pr} % use standard command

\newcommand{\Cmu}       {C_\mu}
\newcommand{\hmu}       {h_\mu}
\newcommand{\EE}        {{\bf E}}

\newcommand{\ProcessAlphabet}   {\MeasAlphabet}

\newcommand{\forward}{+}
\newcommand{\reverse}{-}
\newcommand{\forwardreverse}{\pm} % \pm

\newcommand{\FutureCausalState} { {\CausalState}^{\forward} }

\newcommand{\PastCausalState}   { {\CausalState}^{\reverse} }

\newcommand{\lastindex}[2]{
  \edef\tempa{0}
  \edef\tempb{#2}
  \ifx\tempa\tempb
    \edef\tempc{#1}
  \else
    \edef\tempa{0}
    \edef\tempb{#1}
    \ifx\tempa\tempb
      \edef\tempc{#2}
    \else
      \edef\tempc{#1+#2}
    \fi
  \fi
  \tempc
}

\newcommand{\I}{\mathbf{I}}

\newcommand{\CSjoint}[1][,]{
   \edef\tempa{:}
   \edef\tempb{#1}
   \ifx\tempa\tempb
      \ensuremath{\FutureCausalState\!#1\PastCausalState}
   \else
      \ensuremath{\FutureCausalState#1\PastCausalState}
   \fi
}

\newif\ifpm
\edef\tempa{\forwardreverse}
\edef\tempb{\pm}
\ifx\tempa\tempb
   \pmfalse
\else
   \pmtrue
\fi

\renewcommand{\H}{\operatorname{H}}
\renewcommand{\I}{\operatorname{I}}

\colorlet {R_color}    {blue}
\colorlet {k_color}    {black!30!green}

\def\clap#1{\hbox to 0pt{\hss#1\hss}}

\begin{document}

\title{Structure and Randomness of\\
Continuous-time, Discrete-event Processes}

\author{Sarah E. Marzen}
\email{semarzen@mit.edu}
\affiliation{Physics of Living Systems Group, Department of Physics, Massachusetts Institute of Technology, Cambridge, MA 02139}                     \affiliation{Department of Physics, University of California at Berkeley, Berkeley, CA 94720-5800}

\author{James P. Crutchfield}
\email{chaos@ucdavis.edu}
\affiliation{Complexity Sciences Center, Department of Physics\\
University of California at Davis, One Shields Avenue, Davis, CA 95616}

\date{\today}
\bibliographystyle{unsrt}

% ************************* ABSTRACT *************************
\begin{abstract}
Loosely speaking, the Shannon entropy rate is used to gauge a stochastic
process' intrinsic randomness; the statistical complexity gives the cost of
predicting the process. We calculate, for the first time, the entropy rate and
statistical complexity of stochastic processes generated by finite unifilar
hidden semi-Markov models---memoryful, state-dependent versions of renewal
processes. Calculating these quantities requires introducing novel mathematical
objects (\eMs\ of hidden semi-Markov processes) and new information-theoretic
methods to stochastic processes.
\end{abstract}


\preprint{Santa Fe Institute Working Paper 17-04-XXX}
\preprint{arxiv.org:1704.XXXX [physics.gen-ph]}

\maketitle

% ****************************************************************

%\tableofcontents
\setstretch{1.2}

% Handy abbreviations in the following
\newcommand{\Abet}{\ProcessAlphabet}
\newcommand{\SSet}{\CausalStateSet}
\newcommand{\St}{\CausalState}
\newcommand{\st}{\causalstate}
\newcommand{\MxSt}{\AlternateState}
\newcommand{\MxSSet}{\AlternateStateSet}
\newcommand{\mxst}{\mu}
\newcommand{\mxstt}[1]{\mu_{#1}}
\newcommand{\StartMS}{\bra{\delta_\pi}}
\newcommand{\Ipred}{\EE}
\newcommand{\ISI} { \xi }
\newcommand{\ECT}{\widehat{\EE}}
\newcommand{\CCT}{\widehat{C}_\mu}
\newcommand{\FeatAlphabet}{\mathcal{F}}

%%%%%%%%%%%%%%%%%%%%%%%%%%%%%%%%%%%%%%%%%%

\section{Introduction}

Claude Shannon's seminal $1948$ article ``A Mathematical Theory of
Communication'' introduced a definition for entropy as a well-motivated measure
of randomness \cite{Shan48a}. He further identified entropy rate $\hmu$ as a
measure of the minimal coding cost of a series of potentially correlated
symbols in his celebrated first theorem. In $1989$, Young and Crutchfield
identified statistical complexity $\Cmu$ as the entropy of causal states
\cite{Crut88a}, which are the minimal sufficient statistics for prediction
\cite{Shal98a}. Said simply, $\hmu$ is a measure of a process' intrinsic
randomness and $\Cmu$ a measure of process structure. In one view, these two
measures of complexity are unified by the Kolmogorov-Chaitin complexity of a
discrete object which is the size of the minimal Universal Turing Machine
program that produces the object \cite{Kolm56a,Chai66}. Specifically, the expected
Kolmogorov-Chaitin complexity $\langle K(x_\ell)\rangle$ of a (discrete-time,
discrete-symbol) time series $x_\ell$ of length $\ell$ grows at the Shannon
entropy rate and has an offset determined by the statistical complexity: $\log
\langle K(x_\ell)\rangle \propto_{\ell \to \infty} \Cmu + \ell \hmu$, when
these quantities exist \cite{Crut12a}. Both entropy rate and statistical
complexity have given insight into many disparate complex systems, from chaotic
crystallography~\cite{Varn14a}, biomolecule dynamics
\cite{Kell12a,Li08a,Li13a}, neuronal spike trains \cite{Marz14e}, and animal
behavior \cite{gonzalez2008understanding} to stochastic
resonance~\cite{Witt97a}, geomagnetic volatility~\cite{Clar02a}, hydrodynamic
flows~\cite{Dzug98a,Gonc98a}, and fluid and atmospheric
turbulence~\cite{Palm00a,Cerb13a}.

Perhaps somewhat surprisingly, estimators of the entropy rate and statistical
complexity of \emph{continuous-time}, discrete-symbol processes are lacking.
This is unfortunate since these processes are encountered very often in the
physical, chemical, biological, and social sciences as sequences of discrete
events consisting of an event type and an event duration or magnitude. An
example critical to infrastructure design occurs in the geophysics of crustal
plate tectonics, where the event types are major earthquakes tagged with time
between their occurrence and with an approximate or continuous \emph{Richter
magnitude} \cite{Akim10a}. Understanding these process' randomness and
structure bears directly on averting human suffering. Another example is
revealed in the history of reversals of the earth's geomagnetic field
\cite{Clar03a}, which shields the planet's life from exposure to damaging
radiation. An example from physical chemistry is found in single-molecule
spectroscopy which reveals molecular dynamics as hops between conformational
states that persist for randomly distributed durations \cite{Li13a,Li08a}. The
structure of these conformational transitions is implicated in biomolecular
functioning and so key to life processes. A common example from neuroscience is
found in the spike trains generated by neurons that consist of spike-no-spike
event types separated by continuous \emph{interspike intervals}.  The structure
and randomness of spike trains are key to delineating how tissues support
essential information processing in nervous systems and brains \cite{Marz14e}.
Finally, a growing set of these processes appear in the newly revitalized
quantitative social sciences, in which human communication events and their
durations are monitored as signals of emergent coordination or competition
\cite{Darm13a}.

Here, we provide entropy rate and statistical complexity estimators for
continuous-time, discrete-symbol processes by first identifying the causal
states of stochastic processes generated by continuous-time unifilar hidden
semi-Markov models. In these processes successive dwell times for the symbols
(or events) are drawn based on the process' current ``hidden'' state.
Transitions from hidden state to hidden state then follow a rule that mimics
transitions in discrete-time unifilar hidden Markov models. The resulting
output process consists of the symbols emitted during the dwell time.
Identifying the process causal states leads to new expressions for entropy
rate, generalizing the results of Ref. \cite{Gira05a}, and for statistical
complexity. The hidden semi-Markov process class is sufficiently general that,
in principle, our results yield universal estimators of the entropy rate and
statistical complexity of continuous-time, discrete-event processes.

To start, we define hidden semi-Markov processes and their unifilar generators,
determine their causal states, and use the causal states to calculate their
entropy rate and statistical complexity. We conclude by describing a method for
using the expressions given here to efficiently estimate the entropy rate and
statistical complexity of real-world time series.

\section{Hidden Semi-Markov Processes and Their Unifilar Generators}

\newcommand{\ab}{\allowbreak}

The continuous-time, discrete-symbol process
$\ldots,(\MS_{-1},\MT_{-1}),(\MS_0,\MT_0),(\MS_1,\MT_1),\ldots$ has
realizations $\ldots(\ms_{-1}, \ab \mt_{-1}), \ab (\ms_0,\mt_0),
(\ms_1,\mt_1)\ldots$. \emph{Events} are symbol-duration pairs $(\ms_i,\mt_i)$ that 
occur sequentially in a process.
We demand that $\ms_i\neq\ms_{i+1}$ to
enforce a unique description of the process. In other words, the discrete event symbol
$\ms_i\in\MeasAlphabet$ appears for a total time of $\mt_i$. The present is
located almost surely during the emission of $\ms_0$, and we denote the time
since last emission as $\mt_{0^+}$ and the time to next emission as
$\mt_{0^-}$.  We also, for clarity, denote the last-appearing symbol as
$\ms_{0^+}$ and the next-appearing symbol as $\ms_{0^-}$; though obviously
$\ms_{0^+}=\ms_{0^-}$ almost surely.  (It follows that
$\mt_{0^+}+\mt_{0^-}=\mt_0$.)  The \emph{past}
$\ldots,(\MS_{-1},\MT_{-1}),(\MS_0,\MT_{0^+})$
is denoted $\Past$ and the \emph{future}
$(\MS_0,\MT_{0^-}),(\MS_1,\MT_1),\ldots$ is denoted as $\Future$,
with realizations denoted $\past$ and $\future$, respectively.

A continuous-time, discrete-symbol process' \emph{causal states} $\St$ are the
equivalence classes of pasts defined by the relation:
\begin{align}
\past \sim_\epsilon \pastprime \iff
  \Pr \bigg(\Future \bigg| \Past=\past \bigg)
  = \Pr \bigg(\Future \bigg| \Past=\pastprime \bigg)
  ~.
\label{eq:HSMPEquivReln}
\end{align}
This mimics the relation for the causal states of discrete-time processes
\cite{Shal98a}. A process' \emph{statistical complexity} is the entropy of
these causal states: $\Cmu = \H[\St]$. A process' \emph{prescient states} are
any finer-grained partition of the causal-state classes, as for discrete-time
processes \cite{Shal98a}. Thus, there is a fundamental distinction between a
process' observed or emitted symbols and its internal states. In this way, we
consider general processes as \emph{hidden} processes.

\begin{figure*}
\includegraphics[width=0.49\textwidth]{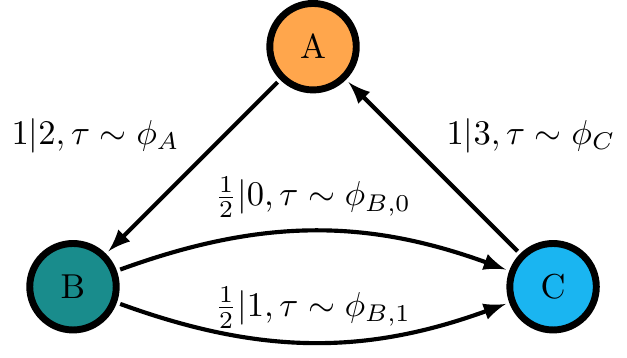}
\caption{Generative model for a unifilar hidden semi-Markov process. The
	notation $p(x)|x,\tau\sim\phi_g$ means that $x$ is emitted with probability
	$p(x)$ from hidden state $g$ with emission time drawn from $\phi_g$.
	}
\label{fig:uHSMMGen}
\end{figure*}

A \emph{hidden semi-Markov process} (HSMP) is a continuous-time,
discrete-symbol process generated by a \emph{hidden semi-Markov model} (HSMM).
A HSMM is described via a hidden-state random variable $\Gen$ with realization
$\gen$, an emission probability $T^{(\ms)}_{\gen}$ for symbol $\ms$, and a
dwell-time distribution $\phi_{\gen}(\tau)$. In other words, in hidden state
$\gen$, symbol $\ms$ is emitted for time $\tau$ drawn from
$\phi_{\gen}(\tau)$.  For reasons that will become clear, we focus on a
restricted form---the \emph{unifilar} HSMM (uHSMM). For these, the present
hidden state $\gen_0$ is uniquely determined by the past emitted symbols
$\ms_{-\infty:0} = \ldots,\ms_{-2},\ms_{-1}$. See Fig.~\ref{fig:uHSMMGen}. In
an abuse of notation, Eq.  (\ref{eq:HSMPEquivReln}) determines a function
$\epsilon(\ms_{-\infty:0})$ that takes the past emitted symbol sequence
$\ms_{-\infty:0}$ to the underlying hidden state $g_0$. (The abuse comes from
suppressing the dependence on durations.) This is the analog appropriate to
this setting of the familiar definition of unifilarity in discrete-time
models---that symbol and state uniquely determine next state.

We assume that the underlying uHSMM has minimal size. That is, out of all such models that generate a given process, we work only with the one having the minimal number of hidden states. Due to the unifilarity constraint, minimality in the number of hidden states is equivalent to minimality in the entropy $\Cmu$ over the hidden states.  A hidden semi-Markov process' minimal uHSMM is its \emph{\eM}.

\section{Causal architecture}

The challenge now is to identify a given HSMP's causal states. With the
causal states in hand, we can calculate their statistical complexity and
entropy rate rather straightforwardly. Theorem \ref{the:HSMPCausalStates} makes the
identification.

{\The A unifilar hidden semi-Markov process' causal states are the triple
$(\gen_0,\ms_{0^+},\tau_{0^+})$, under weak assumptions.
\label{the:HSMPCausalStates}}

{\ProThe To identify causal states, we need to simplify the conditional
probability distribution:
\begin{align*}
p \bigg( \future \bigg| \past \bigg)
  = \Prob \bigg(\Future=\future \bigg| \Past=\past \bigg)
  ~.
\end{align*}
When the process is generated by a unifilar hidden semi-Markov model, this
conditional probability distribution simplifies to:
\begin{align}
p \bigg(\future \bigg|\past \bigg)
  & = p \big((\ms_{0^-},\mt_{0^-}),(\ms_1,\mt_1),\ldots
  \big|\ldots,(\ms_{-1},\mt_{-1}),(\ms_{0^+},\mt_{0^+}) \big)
\nonumber \\
  & = p \big((\ms_1,\mt_1),\ldots \big|\ldots,(\ms_0,\mt_0) \big)
  p \big(\mt_{0^-}
  \big|\ldots,(\ms_{-1},\mt_{-1}),(\ms_{0^+},\mt_{0^+}),\ms_{0^-} \big) \nonumber \\
  & \qquad p \big(\ms_{0^+}
  \big|\ldots,(\ms_{-1},\mt_{-1}),(\ms_{0^+},\mt_{0^+}),\ms_{0^-}) \big)
\label{eq:app1}
\end{align}
where we have $\mt_0 = \mt_{0^+} + \mt_{0^-}$.  Almost surely, we have:
\begin{align}
p \big(\ms_{0^+}\big|
\ldots,(\ms_{-1},\mt_{-1}),(\ms_{0^+},\mt_{0^+}),\ms_{0^-})\big)
  = \delta_{\ms_{0^+},\ms_{0^-}}
\label{eq:app2}
\end{align}
and, due to the unifilarity constraint:
\begin{align}
p \big( \mt_{0^-} \big|
  \ldots,(\ms_{-1},\mt_{-1}),(\ms_{0^+},\mt_{0^+}),\ms_{0^-} \big)
   & = p \big(\mt_{0^-} \big|\mt_{0^+},g_0 = \epsilon^+(\ms_{:0}) \big)
  ~.
\label{eq:app3}
\end{align}
Together Eqs.~(\ref{eq:app1})-(\ref{eq:app3}) imply that the triple $(g_0,\ms_{0^+},\mt_{0^+})$ are prescient statistics. 

These states are causal (minimal and prescient) when two things happen: first,
when the number of hidden states $\gen$ is minimal; and second, when
$\phi_{\gen}(\tau)$ does not take either the ``eventually Poisson'' or
``eventually $\Delta$-Poisson'' form described in Ref. \cite{Marz17a}.  This
last condition is worth spelling out. To avoid an eventually Poisson-like dwell
time distribution, we demand that $\phi_{\gen}(\tau)$ cannot be written as
$\phi_{\gen}(T) e^{-\lambda (t-T)}$ for all $t\geq T$ for some $\lambda>0$ and
$T\geq 0$. To avoid an eventually $\Delta$-Poisson-like dwell time
distribution, we demand that $\phi_{\gen}(\tau)$ cannot be written as:
\begin{align*}
\phi_{\gen}(t) = \phi_{\gen} \big(T+(t-T) \!\!\!\! \mod\Delta \big)
  e^{-\lambda \lfloor (t-T)/\Delta \rfloor}
  ~,
\end{align*}
for any $0<\Delta,\lambda,T<\infty$. Almost all naturally occurring dwell-time
distributions take neither of these forms. And so, we say that the typical
unifilar hidden semi-Markov process' causal states are given by the triple
$(g_0,\ms_{0^+},\mt_{0^+})$.

To find a process' statistical complexity $\Cmu$, we must determine
$H[\St]$ which, in turn, entails finding the probability distribution
$p(\gen_0,\ms_{0^+},\mt_{0^+})$. Implicitly, we are deriving labeled
transition operators as in Ref. \cite{Marz17a}. We start by decomposing:
\begin{align*}
p(\gen_0,\ms_{0^+},\mt_{0^+}) = p(\gen_0) p(\ms_{0^+}|\gen_0) p(\mt_{0^+}|\gen_0,\ms_{0^+})
  ~.
\end{align*}
Since the dwell-time distribution depends only on the hidden state $\gen_0$ and
not on the emitted symbol $\ms_{0^+}$, we find that:
\begin{align*}
p(\mt_{0^+}|\gen_0,\ms_{0^+}) = p(\mt_{0^+}|\gen_0)
  ~.
\end{align*}
As in Ref. \cite{Marz17a}, having a dwell time of at least $\mt_{0^+}$ implies
that:
\begin{align}
p(\mt_{0^+}|\gen_0) & = \int_{\mt_{0^+}}^{\infty} p(\mt_0|\gen_0) d\mt_0
  \nonumber \\
  & = \mu_{\gen_0} \Phi_{\gen_0}(\mt_{0^+})
  ~,
\label{eq:SurvivalDist}
\end{align}
where $\Phi_{\gen_0}(\mt_{0^+})$ will be called the \emph{survival
distribution} and is defined by Eq.~(\ref{eq:SurvivalDist}). From the setup, we
also have:
\begin{align*}
p(\ms_{0^+}|\gen_0) = T_{\gen_0}^{(\ms_{0^+})}
  ~.
\end{align*}

Finally, to calculate $p(\gen_0)$, we consider all ways in which probability can flow from $(\gen',\ms',\tau')$ to $(\gen,\ms,0)$:
\begin{align}
p(\gen,\ms,0)
  & = \sum_{\gen',\ms'} \int_0^{\infty}
  p(\gen',\ms',\tau') p((\gen',\ms',\tau')\rightarrow (\gen,\ms,0)) d\tau'
  ~.
\label{eq:eqdist}
\end{align}
The transition probability $p((\gen',\ms',\tau')\rightarrow (\gen,\ms,0))$ is:
\begin{align}
p((\gen',\ms',\tau')\rightarrow (\gen,\ms,0))
  = \delta_{\gen,\epsilon(\gen',\ms')}
  T_{\gen}^{(\ms)}
  \frac{\phi_{\gen'}(\tau')}{\Phi_{\gen'}(\tau')}
  ~.
\label{eq:transprob}
\end{align}
The term $\delta_{\gen,\epsilon(\gen',\ms')}$ implies that one can only
transition to $\gen$ from $\gen'$ if the emitted symbol $\ms'$ allows. Then,
$T_{\gen}^{(\ms)}$ implies that there is a probability of emitting symbol $\ms$
from newly-transitioned-to hidden state $\gen$. And,
$\phi_{\gen'}(\tau') / \Phi_{\gen'}(\tau')$ is the probability of
emitting $\ms'$ for total time $\tau'$, given that $\ms'$ has already been
emitted for total time at least $\tau'$. Combining Eq.~(\ref{eq:transprob})
with Eq.~(\ref{eq:eqdist}) gives:
\begin{align*}
p(\gen) T_{\gen}^{(\ms)} \mu_{\gen}
  & = \sum_{\gen',\ms'}
  \int_0^{\infty} p(\gen')
  T_{\gen'}^{(\ms')} \mu_{\gen'}
  \Phi_{\gen'}(\tau') \delta_{\gen,\epsilon(\gen',\ms')}
  T_{\gen}^{(\ms)} \frac{\phi_{\gen'}(\tau')}{\Phi_{\gen'}(\tau')} \\
p(\gen)
  & = \frac{1}{\mu_{\gen}}
  \sum_{\gen',\ms'} \mu_{\gen'} p(\gen')
  T_{\gen'}^{(\ms')} \delta_{\gen,\epsilon(\gen',\ms')} \\
  & = \sum_{\gen'} \frac{\mu_{\gen'}}{\mu_{\gen}}
  \left(\sum_{\ms'} T_{\gen'}^{(\ms')}
  \delta_{\gen,\epsilon(\gen',\ms')}\right) p(\gen')
  ~.
\end{align*}
We therefore see that $p(\gen)$ is the eigenvector (appropriately normalized,
$\sum_{\gen} p(\gen) = 1$) associated with eigenvalue $1$ of a transition
matrix given by:
\begin{align}
T_{\gen'\rightarrow \gen}
  := \frac{\mu_{\gen'}}{\mu_{\gen}}
  \left(\sum_{\ms'} T_{\gen'}^{(\ms')}
  \delta_{\gen,\epsilon(\gen',\ms')}\right)
  ~.
\label{eq:TransitionMatrix}
\end{align}
}

\begin{figure*}
\includegraphics[width=0.49\textwidth]{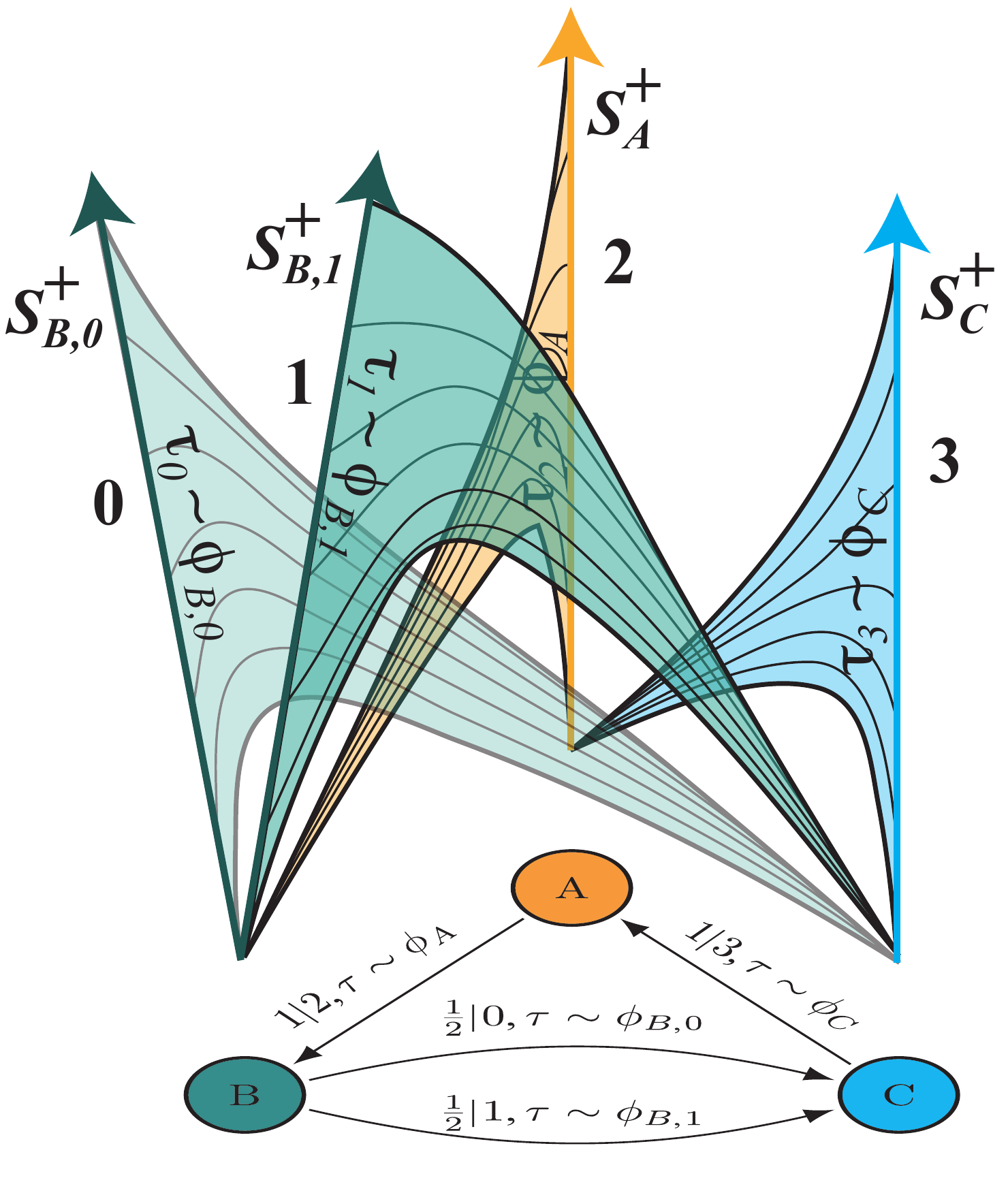}
\caption{Continuous-time \eM\ for the hidden semi-Markov process generated by
	the uHSMM of Figure \ref{fig:uHSMMGen}, as determined by Theorem
	\ref{the:HSMPCausalStates}. Continuous-time causal states $\St_A^+$,
	$\St_B^+$, and $\St_C^+$ track the times $\tau_0$ or $\tau_1$ since last
	event---the times obeying distributions $\phi_A$, $\phi_{B,0}$,
	$\phi_{B,1}$, and $\phi_C$ associated with uHMM states $A$, $B$, and $C$,
	respectively. Each renewal subprocess is depicted as a semi-infinite
	vertical line and is isomorphic with the positive real line.  If no event
	is seen, probability flows towards increasing time since last event, as
	described in Eq. (\ref{eq:TransitionMatrix}). Otherwise, the surfaces
	leaving $\St_A^+$, $\St_B^+$, and $\St_C^+$ indicate allowed transitions
	back to the next reset state or $0$ node located at the non-arrow end,
	denoting that a new event occurred associated with the next state $A$, $B$,
	and $C$, as appropriate. Note that when leaving state $B$ there are two
	distinct diffusion processes on $\St_B^+$ associated with emitting either
	$0$ and $1$. The domains of these diffusions are depicted with two separate
	semi-infinite lines, denoted $\St_{B,0}^+$ and $\St_{B,1}^+$, respectively.
	Figure \ref{fig:uHSMMGen}'s generative HSMM is displayed underneath for
	reference.
	}
\label{fig:uHSMMeM}
\end{figure*}

Drawing from the computer science literature, the \eMs\ of such a process take
on the form of connected counters. See Fig.~\ref{fig:uHSMMeM}.

A process' statistical complexity is defined as the entropy of its causal
states. To calculate this, we need to find the probability distribution over
the triple $(\gen_0,\ms_{0^+},\tau_{0^+})$. After some straightforward
calculation, we find the statistical complexity as stated in
Proposition \ref{prop:Cmu}.

{\Prop The statistical complexity of a unifilar hidden semi-Markov process,
under weak assumptions, is given by:
\begin{align*}
\Cmu = \H[p(\gen)] - \sum_{\gen} p(\gen)
  \left(\sum_{\ms} T^{(\ms)}_{\gen} \log T_{\gen}^{(\ms)}\right)
  - \sum_{\gen} p(\gen) \int_0^{\infty}
  \left(\mu_{\gen} \Phi_{\gen}(\tau)\right)
  \log \left(\mu_{\gen} \Phi_{\gen}(\tau)\right) d\tau
  ~,
\end{align*}
where, as above, $p(\gen)$ is the normalized right eigenvector of eigenvalue
$1$ of the matrix of Eq. (\ref{eq:TransitionMatrix}).
\label{prop:Cmu}
}

{\ProProp Altogether, we find that the statistical complexity is:
\begin{align*}
\Cmu & = \H[\Gen,\MS,\MT] \\
  & = \H[\Gen] + \H[\MS|\Gen] + \H[\MT|\Gen,\MS] \\
  & = \H[p(\gen)] - \sum_{\gen} p(\gen) \left(\sum_{\ms} T^{(\ms)}_{\gen} \log T_{\gen}^{(\ms)}\right) - \sum_{\gen} p(\gen) \int_0^{\infty} \left(\mu_{\gen} \Phi_{\gen}(\tau)\right) \log \left(\mu_{\gen} \Phi_{\gen}(\tau)\right) d\tau,
\end{align*}
where $p(\gen)$ is given above.}

As with continuous-time renewal processes, this is the statistical complexity
of a mixed random variable and, hence, is not always an upper bound on the
excess entropy $\EE := \I \bigg[\Past;\Future\bigg]$.

\section{Informational architecture}
\label{app:information}

Finally, we use the causal-state identification to calculate a uHSMP's entropy
rate. Entropy rates are defined via:
\begin{align*}
\hmu = \lim_{T\rightarrow\infty} \frac{\H \big[\Future^T \big]}{T}
  ~,
\end{align*}
but can be calculated using:
\begin{align*}
\hmu = \lim_{\delta\rightarrow 0} \frac{\H \big[\Future^{\delta}
\big|\Past \big]}{\delta}
  ~.
\end{align*}
Starting there and recalling the definition of causal states, we immediately
have:
\begin{align}
\hmu = \lim_{\delta\rightarrow 0} \frac{\H \big[\Future^{\delta} \big|\St
\big]}{\delta}
  ~.
\label{eq:entrate}
\end{align}
After a little contemplation, we find that entropy rate is as stated in Theorem
\ref{the:EntropyRate}.

{\The The entropy rate of a unifilar hidden semi-Markov process is given by:
\begin{align*}
\hmu = -\sum_{\gen} p(\gen)
  \int_0^{\infty} \mu_{\gen} \phi_{\gen}(\tau)\log \phi_{\gen}(\tau) d\tau
  ~,
\end{align*}
where, as above, $p(\gen)$ is the normalized right eigenvector associated with
eigenvalue $1$ of the matrix in Eq. (\ref{eq:TransitionMatrix}).
\label{the:EntropyRate}
}

{\ProThe As just noted, the entropy rate can be calculated via:
\begin{align*}
\hmu = \lim_{\delta\rightarrow 0} \frac{\H\big[\Future^{\delta} \big|\St
\big]}{\delta}
  ~,
\end{align*}
where $\Future^{\delta}$ are trajectories of time length $\delta$.  Following
Ref. \cite{Marz17a}, we consider the random variable $X_{\delta}$ to be $0$
when emitted symbol is constant throughout the trajectory of length $\delta$,
$1$ when there is one switch from one emitted symbol to another in this
trajectory of length $\delta$, and so on. Basic formulae give:
\begin{align}
\H\big[\Future^{\delta}\big|\St\big] = \H[X_{\delta}|\St] +
\H\big[\Future^{\delta}\big|X_{\delta},\St\big]
  ~.
\label{eq:StateCondFutureEntropy}
\end{align}
The key reason that we condition on $X_{\delta}$ is that two switches are
highly unlikely to happen relative to one switch. Furthermore, if no switches
in emitted symbols occur, then the trajectory is entirely predictable and does
not contribute to the entropy rate. In particular:
\begin{align*}
\Prob(X_{\delta}=1|\St=(g,\ms,\tau))
  & = \int_0^{\delta} \frac{\phi_g(\tau+s)}{\Phi_g(\tau+s)} ds \\
  & \approx \frac{\phi_{\gen}(\tau)}{\Phi_{\gen}(\tau)} \delta
  + O(\delta^2) ~\text{and}\\
  \Prob(X_{\delta}=k|\St=(g,\ms,\tau)) & = O(\delta^k)
\end{align*}
and so:
\begin{align*}
\Prob(X_{\delta}=0|\St=(g,\ms,\tau))
  & = 1-\Prob(X_{\delta}\geq 1|\St=(g,\ms,\tau)) \\
  & = 1 - \frac{\phi_{\gen}(\tau)}{\Phi_{\gen}(\tau)} \delta + O(\delta^2)
  ~.
\end{align*}
In a straightforward way, it follows that:
\begin{align}
\H[X_{\delta}|\St=(g,\ms,\tau)]
  & = -\left(1 - \frac{\phi_{\gen}(\tau)}{\Phi_{\gen}(\tau)}
  \delta\right)\log\left(1 - \frac{\phi_{\gen}(\tau)}{\Phi_{\gen}(\tau)}
  \delta\right) - \left(\frac{\phi_{\gen}(\tau)}{\Phi_{\gen}(\tau)}
  \delta\right) \log \left(\frac{\phi_{\gen}(\tau)}{\Phi_{\gen}(\tau)}
  \delta\right) + O(\delta^2\log\delta) \nonumber \\
  & = \frac{\phi_g(\tau)}{\Phi_{\gen}(\tau)}\delta - \left(\frac{\phi_{\gen}(\tau)}{\Phi_{\gen}(\tau)} \log \frac{\phi_{\gen}(\tau)}{\Phi_{\gen}(\tau)} \right)\delta - \frac{\phi_{\gen}(\tau)}{\Phi_{\gen}(\tau)}\delta\log\delta + O(\delta^2\log\delta)
  ~,
\label{eq:Hdelta1}
\end{align}
after several Taylor approximations; e.g., $\log(1+x)=x + O(x^2)$.

Now, consider the second term in Eq.~(\ref{eq:StateCondFutureEntropy}):
\begin{align*}
\H\big[\Future^{\delta}\big|X_{\delta},\St=(g,\ms,\tau)\big]
  & = \Prob(X_{\delta}=0|\St=(g,\ms,\tau))
  \H\big[\Future^{\delta}\big|X_{\delta}=0,\St=(g,\ms,\tau)\big] \\
  & \qquad + \Prob(X_{\delta}=1|\St=(g,\ms,\tau))
  \H\big[\Future^{\delta}\big|X_{\delta}=1,\St=(g,\ms,\tau)\big] \\
  & \qquad + \sum_{k=2}^{\infty} P(X_{\delta}=k|\St=(g,\ms,\tau))
  \H\big[\Future^{\delta}\big|X_{\delta}=k,\St=(g,\ms,\tau)\big]
  ~.
\end{align*}
If $X_{\delta}=0$, the trajectory is completely determined by $\St=(g,\ms,\tau)$, and hence
\begin{align*}
\H\big[\Future^{\delta}\big|X_{\delta}=0,\St=(g,\ms,\tau)\big] = 0
  ~.
\end{align*}
If $X_{\delta}=1$, then the trajectory is completely determined by one
time---that at which emitted symbols switch. As in Ref. \cite{Marz17a}, the
distribution of switching time is roughly uniform over the interval, and so:
\begin{align*}
\H\big[\Future^{\delta}\big|X_{\delta}=1,\St=(g,\ms,\tau)\big]
  = \log\delta + O(\delta)
  ~.
\end{align*}
Finally, from maximum entropy arguments:
\begin{align*}
P(X_{\delta}=k|\St=(g,\ms,\tau))
  \H\big[\Future^{\delta}\big|X_{\delta}=k,\St=(g,\ms,\tau)\big]
\end{align*}
is at most of $\delta^k (\log\delta)^k$. In particular, we noted earlier that
$P(X_{\delta}=k|\St=(g,\ms,\tau))$ was $O(\delta^k)$ and that $k$ emissions
over a time interval of no more than $\delta$ yields differential entropy of no
more than $(\log\delta)^k$. In addition, we have:
\begin{align*}
\H\big[\Future^{\delta}\big|X_{\delta}= k\big]
  \geq \H\big[\Future^{\delta}\big|X_{\delta}=1\big]
  ~.
\end{align*}
That is, if given a trajectory with a single transition, one can construct
trajectories that approximate it arbitrarily closely with more than one
transition. And so, $\big|\H\big[\Future^{\delta}\big|X_{\delta}= k\big]\big|$
is at least $O(|\log\delta|)$ and at most $O(|\log\delta|^k)$. Hence:
\begin{align*}
\sum_{k=2}^{\infty} P(X_{\delta}=k|\St=(g,\ms,\tau))
  \H\big[\Future^{\delta}\big|X_{\delta}=k,\St=(g,\ms,\tau)\big]
  = O(\delta^2(\log\delta)^2)
  ~.
\end{align*}

Altogether, we have:
\begin{align*}
\H[\Future^{\delta}|\St=(g,\ms,\tau)]
  & = \frac{\phi_g(\tau)}{\Phi_{\gen}(\tau)}\delta - \left(\frac{\phi_{\gen}(\tau)}{\Phi_{\gen}(\tau)} \log \frac{\phi_{\gen}(\tau)}{\Phi_{\gen}(\tau)} \right)\delta - \frac{\phi_{\gen}(\tau)}{\Phi_{\gen}(\tau)}\delta\log\delta + \frac{\phi_{\gen}(\tau)}{\Phi_{\gen}(\tau)} \delta\log\delta + O(\delta^2(\log\delta)^2) \\
  & = \frac{\phi_g(\tau)}{\Phi_{\gen}(\tau)}\delta - \left(\frac{\phi_{\gen}(\tau)}{\Phi_{\gen}(\tau)} \log \frac{\phi_{\gen}(\tau)}{\Phi_{\gen}(\tau)} \right)\delta + O(\delta^2(\log\delta)^2)
\end{align*}
and, thus:
\begin{align*}
\H[\Future^{\delta}|\St]
  & = \left\langle
  \H\big[\Future^{\delta}\big|\St=(g,\ms,\tau)\big] \right\rangle_{g,\ms,\tau}
  \\
  & = \left\langle \frac{\phi_g(\tau)}{\Phi_{\gen}(\tau)}
  - \frac{\phi_{\gen}(\tau)}{\Phi_{\gen}(\tau)}
  \log \frac{\phi_{\gen}(\tau)}{\Phi_{\gen}(\tau)}
  \right\rangle_{g,\ms,\tau} \delta
  + O(\delta^2(\log\delta)^2)
  ~.
\end{align*}

And so, from Eq.~(\ref{eq:entrate}), we find the entropy rate:
\begin{align*}
\hmu & = \lim_{\delta\rightarrow\infty}
  \frac{\H\big[\Future^{\delta}\big|\St\big] }{\delta} \\
  & = \left\langle
  \frac{\phi_g(\tau)}{\Phi_{\gen}(\tau)}
  - \frac{\phi_{\gen}(\tau)}{\Phi_{\gen}(\tau)}
  \log \frac{\phi_{\gen}(\tau)}{\Phi_{\gen}(\tau)}
  \right\rangle_{g,\ms,\tau} \\
  & = \sum_{\gen,\ms} p(\gen) T^{(\ms)}_{\gen} \int_0^{\infty} \mu_{\gen}
  \Phi_{\gen}(\tau)\frac{\phi_g(\tau)}{\Phi_{\gen}(\tau)} d\tau
  - \sum_{\gen,\ms} p(\gen) T^{(\ms)}_{\gen} \int_0^{\infty} \mu_{\gen}
  \Phi_{\gen}(\tau) \frac{\phi_{\gen}(\tau)}{\Phi_{\gen}(\tau)} \log
  \frac{\phi_{\gen}(\tau)}{\Phi_{\gen}(\tau)} \\
  & = \sum_{\gen,\ms} p(\gen) T^{(\ms)}_{\gen}
  \left(\int_0^{\infty} \mu_{\gen} \phi_{\gen}(\tau)d\tau
  - \int_0^{\infty} \mu_{\gen} \phi_{\gen}(\tau)
  \log \frac{\phi_{\gen}(\tau)}{\Phi_{\gen}(\tau)} d\tau \right)
  ~.
\end{align*}
We directly have:
\begin{align*}
\int_0^{\infty} \mu_{\gen} \phi_{\gen}(\tau)d\tau = \mu_{\gen}
\end{align*}
and:
\begin{align*}
\int_0^{\infty} \mu_{\gen} \phi_{\gen}(\tau)
  \log \frac{\phi_{\gen}(\tau)}{\Phi_{\gen}(\tau)}
  & = \int_0^{\infty} \mu_{\gen} \phi_{\gen}(\tau)\log \phi_{\gen}(\tau) d\tau
  - \int_0^{\infty} \mu_{\gen} \phi_{\gen}(\tau)\log \Phi_{\gen}(\tau) d\tau
  ~.
\end{align*}
The second term simplifies substituting $u=\Phi_{\gen}(\tau)$:
\begin{align*}
\int_0^{\infty} \mu_{\gen} \phi_{\gen}(\tau)\log \Phi_{\gen}(\tau) d\tau
  & = - \int_1^0 \mu_{\gen} \log u du \\
  & = \mu_{\gen} \left(u\log u-u\right)|_0^1 \\
  & = -\mu_{\gen}
  ~.
\end{align*}
Altogether, we find:
\begin{align*}
\hmu = -\sum_{\gen} p(\gen)
  \int_0^{\infty} \mu_{\gen} \phi_{\gen}(\tau)\log \phi_{\gen}(\tau) d\tau
  ~.
\end{align*}
}

Theorem \ref{the:EntropyRate} easily generalizes a recent theorem about the
entropy rate of semi-Markov processes \cite{Gira05a} to, essentially, hidden
semi-Markov processes. In other words, Theorem \ref{the:EntropyRate} 
demonstrates the power of causal-state identification.

\section{Conclusion}

Proposition \ref{prop:Cmu} and Theorem \ref{the:EntropyRate} provide new
plug-in estimators for the statistical complexity and entropy rate,
respectively, of continuous-time, discrete-event processes. It might seem that
using these expressions requires an accurate estimation of $\phi_g(\tau)$,
which then might require unreasonable amounts of data. However, a method first
utilized by Lovchenko and popularized by Victor \cite{victor2002binless}
utilizes the fact that we need only calculate scalar functions of
$\phi_g(\tau)$ and not $\phi_g(\tau)$ itself. A second concern arises from the
super-exponential explosion of discrete-time, discrete-alphabet \eMs\ with the
number of hidden states \cite{John10a}. How do we know the underlying topology?
Here, we suggest taking the approach of Ref. \cite{Stre07a}, replacing the
hidden states in these general models with the last $k$ symbols. Further
research is required, though, to determine when the chosen $k$ is too small or
too large.

% \alert{Problem: usually the process is generated by something where there are uncountably infinite counters, but countable counters is dense in the set, so we just keep in mind that we choose a ``nearby'' \eM\ that is justified by some combination of data and expert knowledge.}

\acknowledgments

The authors thank Santa Fe Institute for its hospitality during visits and
thank A. Boyd, C. Hillar, and D. Upper for useful discussions. JPC is an SFI
External Faculty member. This material is based upon work supported by, or in
part by, the U.S. Army Research Laboratory and the U. S. Army Research Office
under contracts W911NF-13-1-0390 and W911NF-12-1-0288. S.E.M. was funded by a
National Science Foundation Graduate Student Research Fellowship, a U.C.
Berkeley Chancellor's Fellowship, and the MIT Physics of Living Systems
Fellowship.

%=================================================================
% References:
%=================================================================
% Use the following option to include external BibTeX files:
\bibliography{chaos}

\end{document}